\documentclass[11pt]{article}
\usepackage{amsmath,amssymb,color,epsfig,cite}
\usepackage{graphicx}
\usepackage{subfigure}
\usepackage{setspace}

\usepackage{amsfonts}

\newcommand{\hoch}[1]{$\, ^{#1}$}

\makeatletter
\@addtoreset{equation}{section}
\makeatother

\newcommand{\be}{\begin{equation}}
\newcommand{\ee}{\end{equation}}
\newcommand{\bea}{\setlength\arraycolsep{2pt} \begin{eqnarray}}
\newcommand{\eea}{\end{eqnarray}}
\newcommand{\nn}{\nonumber}

\def\ft#1#2{{\textstyle{\frac{\scriptstyle #1}{\scriptstyle #2} } }}
\def\fft#1#2{{\frac{#1}{#2}}}

\def\0{{\sst{(0)}}}
\def\1{{\sst{(1)}}}
\def\2{{\sst{(2)}}}
\def\3{{\sst{(3)}}}
\def\4{{\sst{(4)}}}
\def\5{{\sst{(5)}}}
\def\6{{\sst{(6)}}}
\def\7{{\sst{(7)}}}
\def\8{{\sst{(8)}}}
\def\9{{\sst{(9)}}}

\def\sst#1{{\scriptscriptstyle #1}}

\begin{document}



\begin{center}
{\large {\bf Holographic $(a,c)$-charges and Their Universal Relation\\
 in $d=6$ from Massless Higher-order Gravities}}

\vspace{10pt}

H. L\"u\hoch{1\dagger} and Rui Wen\hoch{2*}

\vspace{15pt}

\hoch{1}{\it Center for Joint Quantum Studies, School of Science\\
                  Tianjin University, Tianjin 300350, China}

\vspace{10pt}

\hoch{2}{\it School of Physics, University of Science and Technology of China, Hefei 230026 , China }

\vspace{40pt}

\underline{ABSTRACT}
\end{center}

Recent studies of holographic properties of massless higher-order gravities, whose linear spectrum contains only the (massless) graviton, yielded some universal relations in $d=4$ dimensions between the holographic $a$, $c$ charges and the overall coefficient factor ${\cal C}_T$ of the energy-momentum tensor two-point function, namely $c=\fft1{d-1} \ell\fft{\partial a}{\partial \ell}={\cal C}_T$, where $\ell$ is the AdS radius. The second equality was shown to be valid in all dimensions.  In this paper, we establish the first equality in $d=6$ by examining these quantities from $D=7$ higher-order gravities. Consequently the overall coefficient of the two-point function of the energy-momentum tensor is proportional to this $c$ charge, generalizing the well-known $d=4$ result. We identify the relevant $c$ charge as the coefficient of a specific combination of the three type-B anomalies.  Modulo total derivatives, this combination involves Riemann tensor at most linearly.

\vfill {\footnotesize \hoch{\dagger}mrhonglu@gmail.com\ \ \ \hoch{*}wenrui1024@outlook.com}

\pagebreak

\tableofcontents
\addtocontents{toc}{\protect\setcounter{tocdepth}{2}}



\section{Introduction}
\label{sec:intro}

Owing to the enlarged global symmetry, conformal field theories (CFTs) are much simpler than the general quantum field theories and many tantalizing features have attracted considerable attentions. The subject has been further driven by the advancement of the powerful (anti-de Sitter) AdS/CFT correspondence, which may provide a new window for a deeper understanding of both classical and quantum gravity.  An important feature of any CFT in even $d=2n$ dimensions is that conformal anomalies can arise at the quantum level when the theory is put on a generic curved spacetime background \cite{Duff:1977ay,Duff:1993wm,Imbimbo:1999bj}.  There are two types of conformal anomalies. The type A has the structure of the Euler density and the type B is associated with the Weyl invariants.  In two dimensions, there exists only one type and its constant coefficient is called the central charge or the $c$ charge. In $d=4$ dimensions, there is one central charge associated with either type, and they are called $a$ and $c$ charges respectively.  In higher dimensions there is always one $a$ charge, but the situation for the type-B anomalies becomes increasingly complicated.

The conformal symmetry has the power that many universal properties can be uncovered without having to know the details of a specific theory.  The two-point function of the energy-momentum tensor is completely determined up to an overall coefficient $C_T$ \cite{Osborn:1993cr,Erdmenger:1996yc,Coriano:2012wp}. Furthermore in $d=4$, this coefficient can be established to be proportional to the $c$ charge \cite{Osborn:1993cr,Erdmenger:1996yc}.  This result was reproduced by the holographic technique in Einstein \cite{Liu:1998bu} and Einstein-Gauss-Bonnet \cite{Buchel:2009sk} gravities and more \cite{Myers:2010jv}.  The topic was revisited \cite{Li:2018drw} in the context of higher-order gravities where Einstein gravity with a cosmological constant is extended with higher-order curvature invariant polynomials.  Choosing the parameters appropriately, the theories become massless gravities in that the linear spectrum of the AdS vacuum contains only the usual (massless) graviton. Massless gravities include Einstein, Einstein-Gauss-Bonnet, general Lovelock gravities and more. For general curvature tensor polynomials, there are two massive modes, one scalar and one spin-2. The decoupling of these two modes provides only two constraints on the coupling constants. Therefore massless high-order gravities are abundant. By studying the holography in higher-order massless gravities, a universal relation between $C_T$ and the $a$ charge was uncovered.  In terms of ${\cal C}_T$, which is $C_T$ but with some inessential pure numeric numbers stripped off for a slimmer looking formula,  the relation can be stated as \cite{Li:2018drw}
\be
{\cal C}_T =\fft{1}{d-1} \ell \fft{\partial a}{\partial \ell}\,,\label{identity1}
\ee
where $\ell$ is the AdS radius. (The precise ratio $C_T/{\cal C}_T$ is given in (\ref{ct}).) (This result was also derived in the general off critical case in quasi-topological gravity \cite{Sotkov:2012kx}.) To clarify the equation further, we note that higher-order gravities involve the bare cosmological constant $\Lambda_0$ and $\alpha_i$'s that denote the couplings of higher-order terms.  The equation of motion for an AdS metric of radius $\ell$ implies that
\be
\Lambda_0 = \Lambda_0 (\ell, \alpha_i)\,.
\ee
We can then express ${\cal C}_T={\cal C}_T(\ell, \alpha_i)$ and $a=a(\ell, \alpha_i)$.  In other words, the identity (\ref{identity1}) is valid with the assumption that $(\ell,\alpha_i)$ are independent variables and they are indeed. Since the $a$ charge is related to the universal part of the entanglement entropy \cite{Myers:2010tj}, the identity (\ref{identity1}) in an alternative form was obtained in \cite{Bueno:2018yzo}.

In four dimensions, the identity (\ref{identity1}) implies the relation between $a$ and $c$ charges
\be
c=\ft{1}{3} \ell \fft{\partial a}{\partial \ell}\,.\label{d4ca}
\ee
It should be emphasized that the above holographic relation is the leading-order result of the ``large $N$'' expansion. In superconformal field theories, $a$ and $c$ charges may differ by the subleading correction.  Here the origin of the difference between the $a$ and $c$ charges is the higher-order curvature terms in the bulk, and their CFT implications remain to be investigated.

Whilst the identity (\ref{identity1}) can be established in general dimensions, the generalization of (\ref{d4ca}) to higher dimensions is much less straightforward.  This is because there are multiple possibilities of the type-B anomalies, since the number of Weyl invariants increases profusely as the dimension $d=2n$ increases. There is only one in $d=4$, but there are three in $d=6$.  Two of them are simply the two Weyl cubic terms that have simple higher-dimensional generalizations.  The construction of the third one is highly nontrivial and intense research was involved to get it right \cite{Bonora:1985cq,Deser:1993yx,Erdmenger:1997gy,Bastianelli:2000rs,Bastianelli:2000hi}.

The purpose of this paper is to employ higher-order massless gravities in seven dimensions and to compute their holographic $a$ and three $c_i$ charges. We can then identify the relevant $c$ charge such that the three quantities $(c, a, {\cal C}_T)$ can have universal relations as in the case of $d=4$.  The paper is organized as follows.  In section 2, we review the conformal anomalies in $d=6$ CFTs. We recombine the standard $(I_1,I_2,I_3)$ structures according to the order of Riemann tensor polynomials, and define the corresponding $\hat c_i$ charges.  In section 3, we consider Einstein gravity extended with higher-order derivative terms, up to and including six derivatives. We derive the massless conditions and obtain all the conformal anomalous charges. We are able to identify the relevant $c$ charge and establish the universal relations. We also discuss additional holographic constraints on the coupling constants.  We conclude the paper in section 4.

\section{Conformal anomalies in $d=6$}

We begin with a review of the structure of conformal anomalies in $d=6$.  As was mentioned in the introduction, there are two types, A and B. The type A is associated with the third order Euler density $E_6$, defined by
\bea
-E_6 &=& \fft{6!}{2^3}\,\delta^{\mu_1\cdots \mu_6}_{\nu_1\cdots\nu_6} R^{\nu_1\nu_2}_{\mu_1\mu_2}
R^{\nu_3\nu_4}_{\mu_3\mu_4} R^{\nu_5\nu_6}_{\mu_5\mu_6}\nn\\
&=&R^3 -12 R\,R_{\mu\nu} R^{\mu\nu} + 16 R^{\mu}_{\nu} R^{\nu}_\rho R^{\rho}_\mu + 24 R^{\mu\nu} R^{\rho\sigma} R_{\mu\rho\nu\sigma}+ 3 R R^{\mu\nu\rho\sigma} R_{\mu\nu\rho\sigma}\cr
 &&-24 R^{\mu\nu} R_{\mu \alpha\beta\gamma} R_{\nu}{}^{\alpha\beta\gamma} +
4 R^{\mu\nu}{}_{\rho\sigma} R^{\rho\sigma}{}_{\alpha\beta} R^{\alpha\beta}{}_{\mu\nu}-8 R^\mu{}_\nu{}^\alpha{}_\beta R^\nu{}_\rho{}^\beta{}_\gamma R^{\rho}{}_\mu{}^\gamma{}_{\alpha}\,.
\eea
Note that in literature, the definition of $E_6$ typically involves an additional overall factor $8$. We drop this factor in this paper so that our central charge relations will not be burdened with inessential numerical factors. The type-B anomalies all vanish in the conformally flat backgrounds.  There exists three independent Weyl invariant densities, namely \cite{Bastianelli:2000rs,Bastianelli:2000hi}
\bea
I_1 &=& C_{\mu\rho\sigma\nu} C^{\mu\alpha\beta\nu}
 C_{\alpha}{}^{\rho\sigma}{}_\beta\,,\nn\\
I_2 &=& C_{\mu\nu\rho\sigma} C^{\rho\sigma\alpha\beta}
           C_{\alpha\beta}{}^{\mu\nu}\,,\nn\\
I_3 &=& C_{\mu\rho\sigma\lambda}\Big(\delta^\mu_\nu\, \Box +
    4R^\mu{}_\nu - \ft65 R\, \delta^\mu_\nu\Big) C^{\nu\rho\sigma\lambda}
   + \nabla_\mu J^\mu\,,\\
   \hbox{with}\qquad
J^{\mu}&=&4R_{\mu}^{~\lambda\rho\sigma}\nabla^{\nu}R_{\nu\lambda\rho\sigma}+
3R^{\nu\lambda\rho\sigma}\nabla_{\mu}R_{\nu\lambda\rho\sigma}
-5R^{\nu\lambda}\nabla_{\mu}R_{\nu\lambda}
+\ft12R\nabla_{\mu}R\cr
&&-R_{\mu}^{~\nu}\nabla_{\nu}R+2R^{\nu\lambda}\nabla_{\nu}R_{\lambda\mu}\,.
\eea
It is of interest to note that one can construct a three-parameter family of conformal gravity in six dimensions \cite{mets,Lu:2011ks,Lu:2013hx} with the Lagrangian $\sqrt{-g} (\beta_1 I_1 + \beta_2 I_2 + \beta_3 I_3)$.  For generic parameters $\beta_i$, the theory does not admit the Schwarzschild metric as a solution.  However, there exists one particular combination such that the Schwarzschild metric is indeed a solution. The combination is
\bea
\hat I_1 &=& I_3 - 12 I_1 - 3 I_2\cr
&=& -3 RR^{\mu\nu}R_{\mu\nu}+\ft{9}{25} R^3+6R^{\mu\nu}R^{\rho\sigma}
R_{\mu\rho\nu\sigma} +3R^{\mu\nu}\Box R_{\mu\nu}-\ft{9}{10}R\Box
R\cr
&&+ \fft{1}{\sqrt{-g}}\,\times\,\hbox{total derivatives}\,.
\eea
In other words, modulo total derivative, this combination contains Riemann tensor at most linearly.  Consequently, all Einstein metrics, including the Schwarzschild or the Kerr, are solutions in conformal gravity $\sqrt{-g} \hat I_1$. The most general spherically-symmetric and static black holes in this theory was constructed in \cite{Lu:2013hx}.  This combination is related to the $Q$-curvature, defined by (in our $E_6$ convention)
\be
6Q_6 + I_3 -3 I_2 - 12 I_1 = E_6\,.
\ee
The $Q$-curvature arises typically in supersymmetric CFTs in $d=6$ \cite{Beccaria:2015ypa}, analogous to the simpler example of $Q_4=E_4 - C^2$ that arises in supersymmetric CFTs in $d=4$.

The conformal anomalies in $d=6$ can now be expressed as
\be
(2\pi)^3 \langle T_{\mu}{}^\mu\rangle = - a\, E + c_1 I_1 + c_2 I_2 + c_3 I_3\,,\label{accharges1}
\ee
where the constant coefficients $(a,c_1,c_2,c_3)$ are called the central charges.  The purpose of this paper is not simply to evaluate these central charges using the holographic technique. We shall also use the data to determine whether there is a universal relation between the $a$ charge and the certain combination of the $c$ charges.  We find that it is advantageous to recombine the type-B anomalies, according to the order of the Riemann tensor polynomials.  To be specific, we define
\be
\hat I_1= I_3 - 12 I_1 - 3 I_2\, \qquad \hat I_2= I_2 + 2 I_1\,,\qquad \hat I_3 = I_1\,.
\ee
Modulo totally derivative terms (including $E_6$), the new hatted $\hat I_i$'s involve Riemann polynomials of order $i$.  (In literature, combinations of $I_i$'s based on supersymmetry were discussed in \cite{Butter:2016mtk,Butter:2017jqu,Liu:2017ruz}.)
In terms of these new combinations, the conformal anomalies are
\be
(2\pi)^3 \langle T_{\mu}{}^\mu\rangle = - a\, E + \hat c_1 \hat I_1 + \hat c_2 \hat I_2 + \hat c_3 \hat I_3\,,
\label{accharges2}
\ee
where the $a$ charge remains the same, whilst the type-B charges become
\be
\hat c_1 = c_3\qquad
\hat c_2 = c_2 + 3 c_3\,,\qquad
\hat c_3= c_1 -2 c_2 + 6 c_3\,.\label{hattedcdef}
\ee
It can be seen that if the bulk gravity involves only Ricci polynomials or Riemann tensor at most linearly, then we must have $a=\hat c_1$ and $\hat c_2=0=\hat c_3$. In \cite{Bugini:2016nvn}, the $Q$-curvature was advocated to replace $E_6$, which would shift the $c_i$ charges by the $a$ charge. Our proposal is quite different.

\section{Holographic central charges from higher-order gravities}

In this section, we consider Einstein gravity with a negative cosmological constant, extended with higher order invariant polynomials of curvature tensor and their covariant derivatives, up to and including six derivatives.
We then follow the standard technique to compute the holographic anomalies and derive some universal relations.

\subsection{Higher-order massless gravities}

The Lagrangian of general higher-order gravities, up to and including six derivatives, is
\be
{\cal L}=\sqrt{-g} (L_0 + L_1 + L_2 + L_3 + \tilde L_3)\,.
\ee
where
\bea
L_0&=&-2\Lambda_0\,,\qquad L_1 = R\,,\qquad
L_2 = \alpha_1 R^2 + \alpha_2 R^{\mu\nu} R_{\mu\nu} +
\alpha_3 R^{\mu\nu\rho\sigma} R_{\mu\nu\rho\sigma}\,,\nn\\
L_3&=&\beta_1 R^3 +\beta_2 R\,R_{\mu\nu} R^{\mu\nu} + \beta_3 R^{\mu}_{\nu} R^{\nu}_\rho R^{\rho}_\mu + \beta_4 R^{\mu\nu} R^{\rho\sigma} R_{\mu\rho\nu\sigma}+ \beta_5 R R^{\mu\nu\rho\sigma} R_{\mu\nu\rho\sigma}\cr
 &&+\beta_6 R^{\mu\nu} R_{\mu \alpha\beta\gamma} R_{\nu}{}^{\alpha\beta\gamma} +
 \beta_7 R^{\mu\nu}{}_{\rho\sigma} R^{\rho\sigma}{}_{\alpha\beta} R^{\alpha\beta}{}_{\mu\nu}+\beta_8 R^\mu{}_\nu{}^\alpha{}_\beta R^\nu{}_\rho{}^\beta{}_\gamma R^{\rho}{}_\mu{}^\gamma{}_{\alpha}\,,\nn\\
\tilde L_3 &=& \gamma_1 R\Box R + \gamma_2 R^{\mu\nu}\Box R_{\mu\nu} + \gamma_3 R^{\mu\nu\rho\sigma}
\Box R_{\mu\nu\rho\sigma}\,.
\eea
The theory, or part of it, has been studied for a variety of purposes, see, e.g.~\cite{quasi0,quasi1,Lu:2011zk,Sisman:2011gz,Lu:2015cqa,Bueno:2016xff,Bueno:2016ypa,
Liu:2017kml,Li:2017ncu,Li:2017txk,Feng:2018qnx,Wang:2018neg,Li:2019auk}. (See in particular a recent comprehensive summary of its applications in holography \cite{Li:2019auk} and also \cite{Bueno:2016ypa} for many useful formulae.) It is worth pointing out that there are total 17 six-derivative terms \cite{Bastianelli:2000rs} and six of them are total derivatives.  We use three $\alpha_i$'s and eight $\beta_i$'s to denote the coupling constants of the quadratic and cubic polynomials, and use three $\gamma_i$'s to denote the couplings of those involving explicit derivatives.  For our purpose, we focus on $D=7$ dimensions. The theory admits an AdS vacuum of radius $\ell$, satisfying the on-shell equation
\bea
\fft{15}{\ell_0^2} &=&\fft{15}{\ell^2} - \fft{18}{\ell^4} (21 \alpha_1 + 3\alpha_2 + \alpha_3)\cr
&& +\fft{3}{\ell^6} (1764 \beta _1+252 \beta _2+36 \beta _3+36 \beta _4+84 \beta _5+12 \beta _6+4 \beta _7+5 \beta _8)\,,\label{eom1}
\eea
where we expressed the bare cosmological constant as $\Lambda_0\equiv15/\ell_0^2$. Note that $\gamma_i$'s are absent in the equation.

In literature, it is common to introduce dimensionless couplings $\tilde \alpha_i =\alpha_i/\ell_0^2$, $\tilde \beta_i =\beta_i/\ell_0^4$ and
$\tilde \gamma_i=\gamma_i/\ell_0^4$.  The equation of motion can then be expressed in terms of dimensionless
quantities as
\bea
15 &=&15 f_\infty - 18f_{\infty}^2 (21 \tilde\alpha_1 + 3\tilde\alpha_2 + \tilde\alpha_3)\cr
&& +3f_{\infty}^3 (1764 \tilde \beta _1+252 \tilde \beta _2+36 \tilde \beta _3+36 \tilde \beta _4+84 \tilde \beta _5+12 \tilde \beta _6+4 \tilde \beta _7+5 \tilde \beta _8)\,,\label{eom2}
\eea
where $f_\infty \equiv \ell_0^2/\ell^2$. However, the on-shell condition expressed as our (\ref{eom1}) has the advantage that all the higher-order coupling constants $\alpha_i$, $\beta_i$ and the AdS radius $\ell$ can be viewed as independent variables, with the bare cosmological constant solved as a function of these variables, namely
\be
\Lambda_0 \equiv \fft{15}{\ell_0^2} = \Lambda_0(\ell, \alpha_i, \beta_i)\,.
\ee
In other words, we treat $\Lambda_0$ here as a freely-adjustable parameter. Consequently, all the other holographic quantities such as the conformal charges or the coefficient of two-point functions are also functions of these independent variables $(\ell, \alpha_i, \beta_i)$.  As was discussed in the introduction, this is important for our universal relations since they involve differential operator $\ell \partial/\partial \ell$. This property is lost in the dimensionless constraint (\ref{eom2}) since $(f_\infty, \tilde \alpha_i,\tilde \beta_i)$ cannot be all independent.

We now consider the linear perturbation on the AdS background. Owing to the higher-order derivative terms, the linear spectrum contains two massive scalars and two massive spin-2 modes, in addition to the usual (massless) graviton mode.  We are interested in massless gravity where all these massive modes decouple from the spectrum.  This requires a set of linear constraints on the coupling constants. In particular, the decoupling of the massive scalar modes requires
\bea
&& 24 \gamma_1 + 7 \gamma_2 + 4 \gamma_3=0\,,\cr
&& (24 \alpha_1 + 7 \alpha_2 + 4 \alpha_3) \ell^2 + 10(12\gamma_1 + \gamma_2)\cr
&&-3024 \beta_1 - 582 \beta_2 - 126 \beta_3 - 101 \beta_4 - 264 \beta_5 - 52 \beta_6 - 24 \beta_7 -
15 \beta_8=0\,,\label{massless0}
\eea
and the decoupling of the massive spin-2 modes requires
\bea
&& \gamma_2 + 4 \gamma_3=0\,,\cr
&&(\alpha_2 + 4 \alpha_3) \ell^2 + 1056 \gamma_1 + 659 \gamma_2 + 1564 \gamma_3 \cr
&&- 42 \beta_2 - 18 \beta_3 - 11 \beta_4 - 168 \beta_5 - 28 \beta_6 - 24 \beta_7 + 3 \beta_8=0\,.
\label{massless2}
\eea
Note that one set of massive scalar and spin-2 modes are generated by the $\gamma$-terms entirely. For later purpose, we would like to require that these coupling constants are all independent of $\ell$, which implies that the constraints on $\alpha_i$ should be solved independently from the other constraints.  Thus we have
\bea
&&\{\alpha_1,\alpha_2,\alpha_3\} = \alpha \{1,-4,1\}\,,\qquad
\{\gamma_1,\gamma_2,\gamma_3\}= \gamma \{1,-4,1\}\,,\nn\\
&&\beta_7=-21 \beta_1 - \ft{11}2 \beta_2 - \ft32 \beta_3 - \ft{13}{12} \beta_4 -
\ft{23}3 \beta_5 - \ft43 \beta_6\,,\nn\\
&&\beta_8=-168 \beta_1 - 30 \beta_2 - 6 \beta_3 - 5 \beta_4 - \ft{16}3 \beta_5 - \ft43 \beta_6 +
\ft{16}3 \gamma\,.\label{massless}
\eea
The $\alpha$-terms become the Gauss-Bonnet combination and the $\gamma$-terms become the Gauss-Bonnet inserted with a Laplacian operator. After decoupling all the massive modes, the linear equation involves only the (transverse and traceless) graviton $h_{\mu\nu}$, satisfying
\be
\fft{\kappa_{\rm eff}}{16\pi}\, \big(\bar \Box + \fft{2}{\ell^2}\big)h_{\mu\nu}=0\,,
\ee
where
\be
\kappa_{\rm eff} = 1 - 24 \alpha \ell^{-2} + 4 (126 \beta_1 + 12 \beta_2 + \beta_4 + 4 \beta_5 + 4 \gamma)\ell^{-4}\,.\label{keff}
\ee
Note that we have normalized our convention such that $\kappa_{\rm eff}=1$ for Einstein gravity. If we turn off all the quadratic or higher-order Riemann polynomials, the massless gravity becomes that of Ricci-polynomial quasi-topological with $\kappa_{\rm eff}=1$ \cite{Li:2017ncu}. It is of interest to note that
\be
\kappa_{\rm eff} = \fft{\ell^3}{30} \fft{\partial \Lambda_0}{\partial \ell}\,.
\ee
The equivalent form of this identity, with the differentiation done by $f_\infty$, was observed and proven in \cite{Bueno:2018yzo}.

According to the AdS/CFT dictionary, the massless graviton is associated with the boundary stress tensor. Following the standard procedure, one can obtain its two-point function.  The overall coefficient is then clearly proportional to $\kappa_{\rm eff}$, since this is the only parameter left in the linearized equation.  It turns out one has for general $d$ dimensions \cite{Li:2018drw}
\be
C_T = \fft{\Gamma(d+2)}{8 (-1)^{\fft{d}{2}}(\pi)^{\fft{d}2 +1} (d-1)\Gamma(\fft{d}2)} {\cal C}_T\,,\qquad {\cal C}_T=\kappa_{\rm eff} \ell^{d-1}\,.\label{ct}
\ee

\subsection{Holographic anomalous charges}

For given gravity with an AdS vacuum in odd $D=d+1$ dimensions, there is a standard method to compute the holographic anomaly \cite{Henningson:1998gx,Skenderis:2002wp}, using the Fefferman-Graham (FG) expansion around the AdS boundary. Substituting the FG ansatz into the Lagrangian, one finds that there exists a divergent term that cannot be cancelled by the boundary terms such as the Gibbons-Hawking surface term and/or the holographic counterterms. Requiring that the ansatz satisfy the equations of motion, one can then equate the anomalous term to (\ref{accharges1}) (or equivalently (\ref{accharges2})) and read off the coefficients $a$ and $c_i$. For a generic boundary metric, this procedure can be very involved for higher-order gravities.  However, if one is simply to read off the central charges, one can choose some specific boundary ans\"atze with isometries such as $S^6$, $S^4\times S^2$, $S^2\times S^2\times S^2$, etc.  The central charges were computed in \cite{Bugini:2016nvn}.  In our convention, we find
\bea
a &=&\ell^5  - 4 (21 \alpha_1 + 3 \alpha_2 + \alpha_3) \ell^3\cr
&& +3 (1764 \beta_1 + 252 \beta_2 + 36 \beta_3 + 36 \beta_4 + 84 \beta_5 + 12 \beta_6 + 4 \beta_7 +
    5 \beta_8) \ell\,,\cr
c_1 &=& -12 \ell^5 + 16 (63 \alpha_1 + 9 \alpha_2 - \alpha_3) \ell^3\cr
&& +12 (128 \gamma_3 - 5292 \beta_1 - 756 \beta_2 - 108 \beta_3 - 108 \beta_4 - 28 \beta_5 -
   4 \beta_6 + 20\beta_7 - 39 \beta_8)\ell\,,\cr
c_2 &=&-3\ell^5 +4 (63 \alpha_1 + 9 \alpha_2 + 7 \alpha_3) \ell^3\cr
&&-3 (128 \gamma_3 + 5292 \beta_1 + 756 \beta_2 + 108 \beta_3 + 108 \beta_4 + 476 \beta_5 +
   68 \beta_6 - 20 \beta_7 + 7 \beta_8) \ell\,,\cr
c_3 &=&\ell^5-12 (7 \alpha_1 + \alpha_2 - \alpha_3) \ell^3\cr
&&-3 (64 \gamma_3 - 1764 \beta_1 - 252 \beta_2 - 36 \beta_3 - 36 \beta_4 + 140 \beta_5 +
   20 \beta_6 + 28 \beta_7 - 13 \beta_8) \ell\,.\label{accharges}
\eea
The result for Einstein gravity was first obtained in \cite{Henningson:1998gx}. Note that we followed the normalization convention of \cite{Li:2018drw} such that for Einstein gravity the relevant quantities are given by
\be
a^{\rm Ein}=\ell^{d-1}\,,\qquad \kappa_{\rm eff}^{\rm Ein}=1\,,\qquad {\cal C}_T^{\rm Ein}=\ell^{d-1}\,.
\label{convention}
\ee
It should be emphasized that holographic charges (\ref{accharges}) are obtained from the on-shell action, and therefore the equation of motion (\ref{eom1}) is imposed.  The hatted type-B charges defined by (\ref{hattedcdef}) are given by
\bea
\hat c_1 &=& c_3\,,\cr
\hat c_2 &=&64 \alpha_3 \ell^3 -96 (10 \gamma_3 + 28 \beta_5 + 4 \beta_6 + 2 \beta_7 - \beta_8) \
\ell\,,\cr
\hat c_3 &=& 192 (6 \gamma_3 - 2 \beta_7 - \beta_8) \ell\,.
\eea
We see a hierarchical structure of the hatted charges depending on the order of the Riemann polynomials in the bulk Lagrangian. In particular, in Einstein or Ricci-polynomial gravities, both $\hat c_2$ and $\hat c_3$ vanish, with only $\hat c_1$ non-vanishing and it equals to $a$. This property was commented under (\ref{hattedcdef}).

The higher-order gravities become all massless once we impose the conditions (\ref{massless}), in which case, we find
\bea
\hat c_1 &=& c_3= \ell^5 -24 \alpha \ell^3 + 4  (126 \beta_1 + 12 \beta_2 + \beta_4 + 4 \beta_5 + 4 \gamma)\ell\,,\cr
a &=& \ell^5 -40 \alpha \ell^3 + 20 (126 \beta_1 + 12 \beta_2 + \beta_4 + 4 \beta_5 + 4 \gamma)\ell\,.
\eea
Thus it is natural to single out $\hat c_1$ and give a new name $c=\hat c_1$.  We find the universal relation
\be
c=\fft{1}{5}\, \ell\fft{\partial a}{\partial\ell}\,.
\ee
This generalizes the four-dimensional result (\ref{d4ca}).  It follows from (\ref{keff}) and (\ref{ct}) that
\be
{\cal C}_T=c\,.
\ee
In literature, the holographic charges are typically expressed in terms of dimensionless parameters. For simplicity, we consider the Einstein-Gauss-Bonnet theory as an example.  The holographic quantities are
\be
C_{T} \sim c\sim \ell_0^5 f_{\infty}^{-\fft52} \Big(1 - 24 \tilde \alpha f_\infty\Big)\,,\qquad
a\sim  \ell_0^5 f_{\infty}^{-\fft52} \Big(1 - 40 \tilde\alpha f_{\infty}\Big)\,.
\ee
Thus one can propose a universal identity \cite{Bueno:2018yzo}
\be
C_T \sim f_\infty \fft{\partial a}{\partial f_{\infty}}\,.
\ee
To be precise, $a$ was the free energy of the squashed sphere in \cite{Bueno:2018yzo}. This equality is true only if one treats $\tilde \alpha$ as being independent of $f_\infty$ and it hence requires one to ignore the on-shell condition $12\tilde \alpha f_\infty^2 - f_\infty +1=0$.  In our parametrization, on the other hand, with the help of the freely-adjustable bare cosmological constant $\Lambda_0=\Lambda_0(\ell, \alpha)$, $(\ell, \alpha)$ are indeed independent variables.

\subsection{Further constraints from the holographic $a$-theorem}

For a well-defined CFT, one may expect that there exists an $a$-theorem. It states that the $a$ charge, which measures the massless degrees of freedom, increases monotonically as a function of some RG scale $\mu$. Whilst the theorem was well established in $d=2$ \cite{zamo}, its higher-dimensional generalizations are less concrete
\cite{Cardy:1988cwa,Komargodski:2011vj,Cordova:2015vwa,Cordova:2015fha,Huang:2015sla,Chen:2015hpa}. Holographic $a$-theorems on the other hand are much easier to establish since they are simply related to the null-energy condition of minimally coupled matter in the AdS domain wall metric \cite{Freedman:1999gp,Myers:2010xs,Sahakian:1999bd,Anber:2008js,Myers:2010tj}. An $a$-theorem in Einstein-Horndeski gravity with non-minimally coupled matter was also established at the critical point \cite{Li:2018kqp}.
Technically, one is expected to derive the full equations of motion $T_{\mu\nu}^{\rm mat}=E_{\mu\nu} \equiv \partial {\cal L}/\partial g^{\mu\nu}$ even though the null energy-condition only requires to know $(E_t^t, E_r^r$) in the domain wall metric such that $-E_t^t + E_r^r\ge 0$.  Obtaining $E_{\mu\nu}$ can be involved when the theory becomes complicated.  In \cite{Li:2017txk,Li:2018kqp}, this step was circumvented by introducing a free massless scalar.  This method was also adopted in \cite{Alkac:2018whk} to establish the $c$-theorem in Born-Infeld gravity theories.

We follow the technique of \cite{Li:2017txk,Li:2018kqp}, and find as in \cite{Li:2017txk}, the existence of an $a$-theorem requires the decoupling of the scalar modes, but does not discriminate the ghost modes.  In other words, we find that the coupling constants should satisfy (\ref{massless0}).  In addition, two more constraints should be further imposed, namely,
\bea
&&24\alpha_1 + 7 \alpha_2 + 4 \alpha_3=0\,,\cr
&&288 \beta_1 + 84 \beta_2 + 37 \beta_3 + 12 \beta_4 + 48 \beta_5 + 14 \beta_6 + 8 \beta_7=40 (9 \gamma_1 + 2 \gamma_2 + \gamma_3)\,,
\eea
Together with the ghost free condition (\ref{massless2}), we have
\be
\beta_6 = 24 \gamma_1 - 36 \beta_1 - 12 \beta_2 - \ft{15}{2} \beta_3 - \beta_4 + 4 \beta_5\,.
\ee
The resulting theory includes Einstein-Gauss-Bonnet, Einstein-Lovelock, quasi-topological \cite{quasi0,quasi1} and quasi-topological Ricci \cite{Li:2017ncu,Li:2017txk} gravities.

\section{Conclusions}

Massless gravities are those whose linearized spectrum in a maximally-symmetric background consists only of the graviton. Massless higher-order gravities provide a class of bulk theories for studying the holographic properties, where conformal anomalies, two-point functions and $a$-theorems can be all straightforwardly computed. They are abundant and thus very useful to uncover some universal properties of CFTs. One such an example is the differential identity (\ref{identity1}). In $d=4$, the differential relation (\ref{d4ca}) was also shown, and it is consistent with ${\cal C}_T=c$ that was established in CFTs. In this paper, by computing the holographic conformal anomalies in $D=7$ massless high-order gravities, we found that the
relations
\be
{\cal C}_T = c = \fft15\ell \fft{\partial a}{\partial \ell}
\ee
held in $d=6$. Note that our simple forms of the relations are based on the convention (\ref{convention}).
This result is much more nontrivial than the one in $d=4$ since there are three $c_i$ charges in six dimensions, associated with the three Weyl invariants. We identified that the relevant $c$ charge was associated with the Weyl invariant combination that has at most the linear Riemann tensor, modulo some total derivatives (including the Euler density.) Although in $d=6$, the relevant $c$ is simply the $c_3$, we believe there is an advantage to organize the Weyl invariants according to the order of Riemann tensor polynomials. This organizing principle can be generalized to all dimensions and the relevant $c$ charge is the coefficient of the Weyl invariants with the least Riemann tensor, modulo total derivatives. Alternatively, since $c=c_3$, a more straightforward organization is that the relevant conformal combination takes the form $C\Box^{d-4}  C + \cdots$, in other words; it involves the most explicit covariant derivatives. We expect that the two description of the relevant conformal invariants are equivalent and there should exist only one such $c$ and that ${\cal C}_T=c$ continues to hold in higher dimensions.

The holographic approach using massless higher-order gravities provides enormous insights to the properties of CFTs. It is of great interest to investigate whether these universal relations can be established by CFTs themselves.  The differential relations however present an immediate challenge since they involve $\ell$, whose CFT correspondence is not generally clear.  On the other hand, the relation ${\cal C}_T=c$ involves only the CFT quantities and it was proven in $d=4$. Our holographic conclusion encourages the possibility that this may be proven in six-dimensional CFTs .

\section*{Acknowledgement}

We are grateful to Yue-Zhou Li for useful discussions.  The work is supported in part by NSFC grants No.~11875200 and No.~11475024.

\end{document}